\documentclass[twoside]{dis08}
\usepackage[utf8]{inputenc}
\usepackage[dvips]{graphicx,epsfig,color}
\usepackage{wrapfig,rotating}
\usepackage{amssymb,amsmath,array}

\begin{document}
\title{Azimuthal asymmetries in Deeply Virtual Compton Scattering on
an unpolarized proton target}

%***********************************************************************
% AUTHORS INFORMATION AREA
%***********************************************************************
\author{Dietmar Zeiler on behalf of the HERMES collaboration
%
% Optional short acknowledgment: remove next line if non-needed
%\thanks{This is an optional funding source acknowledgment.}
%
% DO NOT MODIFY THE FOLLOWING '\vspace' ARGUMENT
\vspace{.3cm}\\
%
% Addresses and institutions (remove "1- " in case of a single institution)
University of Erlangen-Nürnberg - Physikalisches Institut II \\
%Erwin-Rommel-Strasse 1, 91058 
Erlangen - Germany
%
% Remove the next three lines in case of a single institution
%\vspace{.1cm}\\
%2- School of Second Author - Dept of Second Author \\
%Address of Second Author's school - Country of Second Author's school\\
}
%***********************************************************************
% END OF AUTHORS INFORMATION AREA
%***********************************************************************

\maketitle

\begin{abstract}
Preliminary results on azimuthal asymmetries in leptoproduction of real
photons on an unpolarized hydrogen target measured at the HERMES experiment
are presented \cite{urlDIS08}. The analysis 
of data taken with different beam charges and helicities allows for a
simultaneous extraction of asymmetries originating from the interference
term and the squared DVCS amplitude. Sizeable asymmetry amplitudes for 
the first cosine moment of the beam-charge asymmetry and the first sine
moment of the beam-spin asymmetry have been found. The results are compared
to different theoretical calculations.
%Place your abstract here. It should not exceed 100 words.
\end{abstract}

\section{Introduction}

Hard exclusive leptoproduction of real photons on nucleons (Deeply Virtual 
Compton Scattering, DVCS) can be described using Generalized 
Parton Distributions (GPDs) \cite{Dittes:1988xz,Muller:1998fv}. These subsume
both Parton Distribution Functions and Form Factors and 
thus provide a 3-dimensional picture of the nucleon.
They also offer a way to access the total angular momentum carried by the
partons in the nucleon \cite{Ji:1996nm}. 

The DVCS process interferes with the Bethe-Heitler (BH) process, because both 
have the same final state consisting of the scattered beam lepton, the recoiling
proton and a real photon. The latter is radiated from the struck quark in the 
case of DVCS and from the lepton in the BH process. This leads to three amplitudes 
squared entering into 
the cross section for exclusive leptoproduction of real photons, namely the 
squared DVCS amplitude $\mathcal{j} \mathcal{T}_{\text{DVCS}} \mathcal{j}^2$, 
the squared BH amplitude $\mathcal{j} \mathcal{T}_{\text{BH}} \mathcal{j}^2$ 
and an interference term $\mathcal{I}$ \cite{Belitsky:2001ns,Diehl:1997bu}:
\begin{equation}
\frac{d \sigma}{dx_B dQ^2 dt d\phi} = \frac{\alpha^3 x_B y}{16  \pi^2 Q^2 e^3} 
\frac{2 \pi y}{Q^2}\frac{\mathcal{j} \mathcal{T}_{\text{DVCS}} \mathcal{j}^2
+ \mathcal{j} \mathcal{T}_{\text{BH}} \mathcal{j}^2 + \mathcal{I}}
{\sqrt{1+ 4 x_B^2 M^2_N / Q^2}}. \nonumber
\end{equation}
The kinematic quantities in this equation are the Bjorken scaling variable
$x_{\text{B}}$, the squared four-momentum transfer mediated by the virtual photon $Q^2$,
the squared four-momentum transfer to the nucleon $t$ and the azimuthal angle 
$\phi $, defined as the angle between the lepton scattering plane and the photon 
production plane. The BH amplitude gives the largest contribution 
at the HERMES kinematics and is calculable in Quantum Electrodynamics. 
However, the extraction of different asymmetries with respect to the
beam charge and/or beam helicity allows to access the suppressed terms. 

The cross section for a longitudinally polarized lepton beam scattered off an 
unpolarized proton target $\sigma_{\text{LU}}$ can be related to the unpolarized
cross section $\sigma_{\text{UU}}$ by:
\begin{equation}
\sigma_{\text{LU}} (\phi;P_{\text{l}},e_{\text{l}}) 
= \sigma_{\text{UU}}(\phi) \cdot \left\{ 1 + P_{\text{l}} A_{\text{LU}}^{\text{DVCS}}(\phi) 
+ e_{\text{l}} P_{\text{l}} A_{\text{LU}}^{\mathcal{I}}(\phi) 
+ e_{\text{l}}  A_{\text{C}} (\phi)
\right\}, \nonumber
\end{equation}
where $e_{\text{l}} (P_{\text{l}})$ denotes the beam charge (polarization). This defines
the charge (in)dependent beam spin asymmetry (BSA) $A_{\text{LU}}^{\mathcal{I}}$ 
($A_{\text{LU}}^{\text{DVCS}}$) and the beam charge asymmetry (BCA) $A_{\text{C}}$. 
In the analysis only effective asymmetry amplitudes can 
be extracted, which include $\phi$-dependencies from the BH propagators and the
unpolarized cross section. 

The above defined asymmetries have been expanded in $\phi$:
\begin{eqnarray}
A_{\text{C}} &=& c_{0,\text{C}} + s_{1,\text{C}} sin \phi + 
c_{1,\text{C}} cos(\phi) + c_{2,\text{C}} cos(2 \phi) + c_{3,\text{C}} cos(3 \phi), \nonumber \\
A_{\text{LU}}^{\text{DVCS}} &=& c_{0,\text{LU}}^{\text{DVCS}} + s_{1,\text{LU}}^{\text{DVCS}} sin \phi 
+ c_{1,\text{LU}}^{\text{DVCS}} cos(\phi) + s_{2,\text{LU}}^{\text{DVCS}} sin(2 \phi), \nonumber  \\
A_{\text{LU}}^{\mathcal{I}} &=& c_{0,\text{LU}}^{\mathcal{I}} + s_{1,\text{LU}}^{\mathcal{I}} sin \phi 
+ c_{1,\text{LU}}^{\mathcal{I}} cos(\phi) + s_{2,\text{LU}}^{\mathcal{I}} sin(2 \phi), \nonumber
\end{eqnarray}
where the $sin \phi$ ($cos \phi$) term in the BCA (BSAs) and the $sin 2\phi$ term in the 
charge-independent BSA have been added as a consistency check. By combining the data 
taken with different beam charges and helicities, the amplitudes have been fit 
simultaneously using a Maximum Likelihood method, which is described in detail 
in \cite{:2008jga}.

\section{Data analysis}

The data has been taken at the HERMES experiment \cite{Ackerstaff:1998av} 
located at the HERA storage ring at DESY. The longitudinally polarized electron 
(positron) beam of $27.6$ GeV energy was used to scatter off a polarized or unpolarized
internal hydrogen gas target. Exclusive events have been identified by requiring the 
detection of exactly one lepton with the same charge as the beam and $Q^2 > 1$ GeV$^2$ and of 
exactly one photon. 
In addition, as the recoiling proton has not been detected, the missing mass $M_{\text{x}}$ 
was required to match the proton mass within the resolution of the spectrometer. With 
these cuts it is not possible to distinguish the elastic DVCS/BH events from the associated 
processes, where the nucleon in the final state is excited to a resonant state. Monte
Carlo (MC) simulations estimate its contribution to about $12 \%$, which is taken as part 
of the signal. The main background contribution with about $3 \%$ is originating from the 
semi-inclusive $\pi^0$ production and is corrected for. The exclusive $\pi^0$ production is
estimated to be less than $0.5 \%$.

The systematic uncertainties are obtained from a MC simulation estimating the effects
of limited acceptance, smearing, finite bin-width and the alignment of the detectors 
with respect to the beam. Other sources are the 
background corrections and a shift of the position of the exclusive missing mass peak 
between the data taken with different beam charges. 

\section{Results}

\begin{figure}[hbt]
\begin{center}
\includegraphics[width=1.\columnwidth]{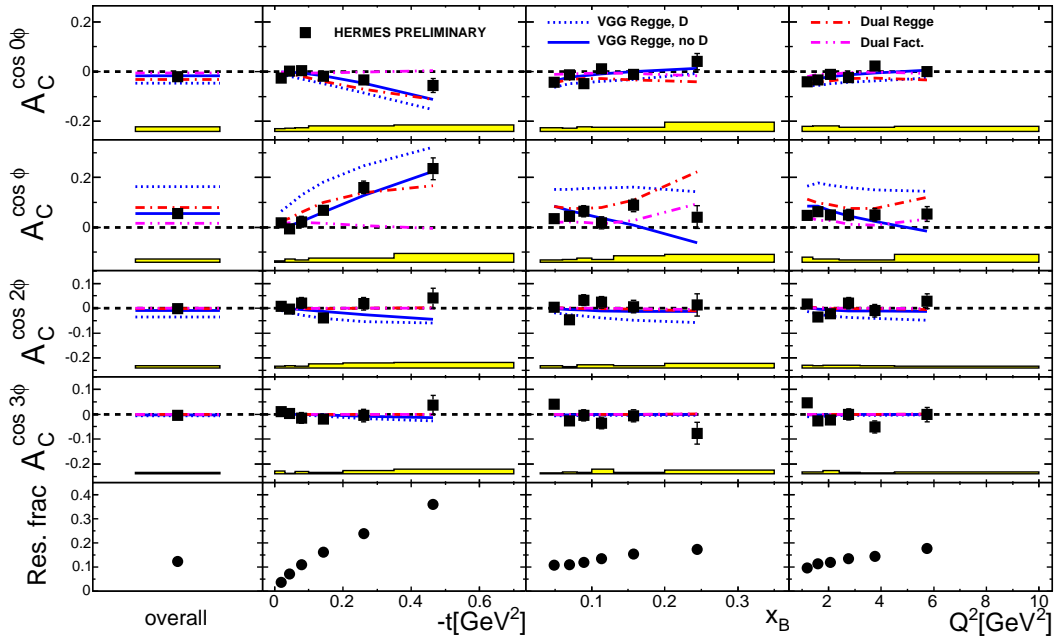}
\end{center}
\vspace{-1.5cm}
\caption[REL-0]
{The BCA amplitudes extracted on a hydrogen target at the HERMES experiment. For more
 explanations see the text.
\label{Bca}}
\vspace{-0.3cm}
\end{figure}

The first four rows of figure \ref{Bca} represent different cosine amplitudes of the 
BCA, whereas the last row displays the fractional contributions of the associated
process. In the first column the integrated result is shown, in the other columns the amplitudes 
are binned in $-t$, $x_{\text{B}}$ and $Q^2$. The error bars represent the statistical
and the bands the systematic uncertainty. The magnitudes of the first two cosine moments 
$A_{\text{C}}^{cos 0\phi}$ and $A_{\text{C}}^{cos \phi}$ increase with increasing $-t$, while having
opposite signs as theoretically expected. Both relate in HERMES kinematics to the real
part of the GPD $H$, but the constant term is suppressed relative to the first moment. The 
second cosine moment appears in twist-three approximation and is found to be compatible with zero
like the third cosine moment, which is related to gluonic GPDs. 
%The 
%fractional contribution of the associated process rises up to $35 \%$ in the last $t$-bin.
 
\begin{figure}[hbt]
\begin{center}
\includegraphics[width=1.\columnwidth]{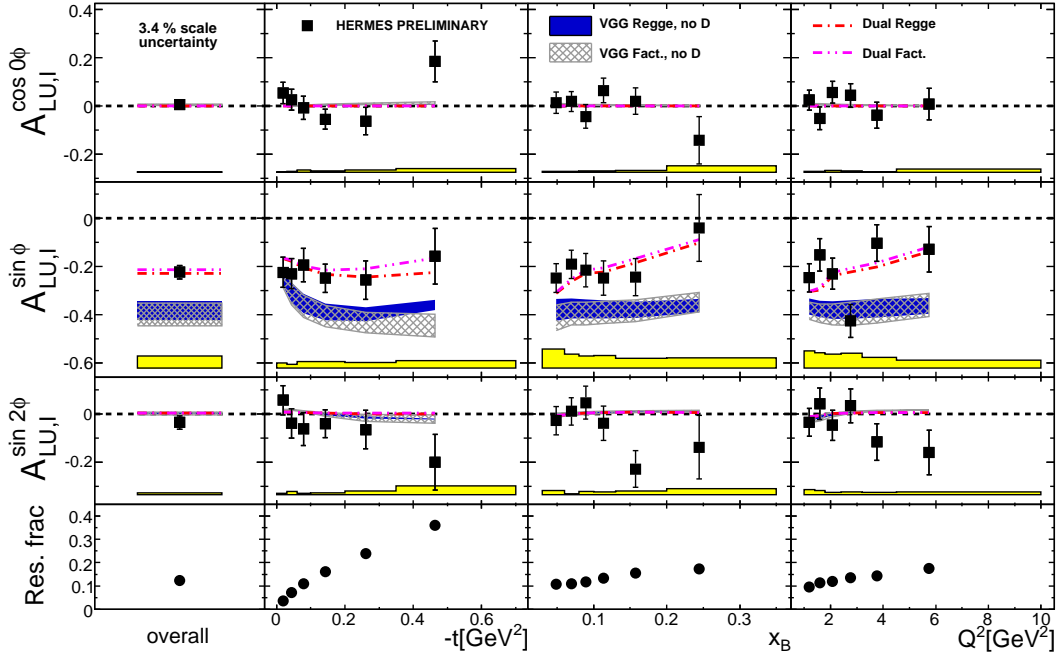}
\end{center}
\vspace{-0.8cm}
\caption[REL-2]
{The charge-dependent BSA amplitudes extracted on a hydrogen target at the HERMES 
experiment. For explanations see the text.
\label{Bsai}}
\vspace{-0.3cm}
\end{figure}

The charge-independent BSA moments are found to be compatible with zero (not shown, see 
\cite{urlDIS08}). On the contrary, the first sine moment $A_{\text{LU,I}}^{sin \phi}$ is
large and negative in the covered kinematics (see figure \ref{Bsai}). This amplitude 
relates to the imaginary part of the GPD $H$. 
The constant moment of the BSA $A_{\text{LU,I}}^{cos 0\phi}$ is compatible with zero, 
while the second sine moment $A_{\text{LU,I}}^{sin 2\phi}$ seems to tend towards 
negative values. 

Also drawn in the figures are model calculations based on the framework of 
double distributions \cite{Vanderhaeghen:1999xj,Goeke:2001tz} (labeled VGG) and the 
dual-parametrization model of \cite{Polyakov:2002wz,Guzey:2006xi} (labeled Dual). 
Both models incorporate a Regge-inspired $t$-ansatz and a factorized $t$-ansatz. 

The BCA amplitudes favor the models with a Regge-inspired 
$t$-dependence, if the D-term is neglected in the case of the DD-based model.
For the comparison of this model to the BSA results bands have been calculated by
varying the skewness dependence via the parameters $b_{\text{val}}$ and  $b_{\text{sea}}$
. Irrespective
of the chosen $t$-ansatz both bands miss to describe the data except for small$-t$. 
However, with increasing $-t$ the unknown contribution from the associated process 
rises up to about $35 \%$.
The dual-parametrization model describes this data very well.
 
To conclude, HERMES has measured a significant first cosine (sine) moment in the BCA 
(charge-dependent BSA) and thereby confirmed previous publications 
\cite{Airapetian:2001yk,Airapetian:2006zr,:2008jga}. The statistical
precision of the data allows for strong constraints on theoretical calculations, once
the influence of the associated process has been pinned down. This will be possible with
the data taken with the Recoil Detector. 

%\section{Acknowledgments}
%\vspace{0.2cm}

This work has been supported by the German Bundesministerium für Bildung und Forschung
(contract nr. 06 ER 143) and the European Community-Research Infrastructure Activity 
(HadronPhysics I3, contract nr. RII3-CT-2004-506078). 

% ****************************************************************************
% BIBLIOGRAPHY AREA
% ****************************************************************************

\begin{footnotesize}
% IF YOU DO NOT USE BIBTEX, USE THE FOLLOWING SAMPLE SCHEME FOR THE REFERENCES
% ----------------------------------------------------------------------------

% IF YOU USE BIBTEX,
% - DELETE THE TEXT BETWEEN THE TWO ABOVE DASHED LINES
% - UNCOMMENT THE NEXT TWO LINES AND REPLACE 'Name_Of_Your_BibFile'

\bibliographystyle{hunsrt}
\bibliography{refs.bib}

\begin{thebibliography}{10}

\bibitem{urlDIS08}
Slides:~\\ {\tt
  http://indico.cern.ch/contributionDisplay.py?contribId=290\&sessionId=31\&co%
nfId=24657}.

\bibitem{Dittes:1988xz}
F.~M. Dittes, D.~Muller, D.~Robaschik, B.~Geyer, and J.~Horejsi.
\newblock The altarelli-parisi kernel as asymptotic limit of an extended
  brodsky-lepage kernel.
\newblock {\em Phys. Lett.}, B209:325, 1988.

\bibitem{Muller:1998fv}
D.~Muller, D.~Robaschik, B.~Geyer, F.~M. Dittes, and J.~Horejsi.
\newblock Wave functions, evolution equations and evolution kernels from
  light-ray operators of {QCD}.
\newblock {\em Fortschr. Phys.}, 42:101, 1994, hep-ph/9812448.

\bibitem{Ji:1996nm}
X.~Ji.
\newblock Deeply-virtual compton scattering.
\newblock {\em Phys. Rev.}, D55:7114, 1997, hep-ph/9609381.

\bibitem{Belitsky:2001ns}
Andrei~V. Belitsky, Dieter Mueller, and A.~Kirchner.
\newblock {Theory of deeply virtual Compton scattering on the nucleon}.
\newblock {\em Nucl. Phys.}, B629:323--392, 2002, hep-ph/0112108.

\bibitem{Diehl:1997bu}
Markus Diehl, Thierry Gousset, Bernard Pire, and John~P. Ralston.
\newblock {Testing the handbag contribution to exclusive virtual Compton
  scattering}.
\newblock {\em Phys. Lett.}, B411:193--202, 1997, hep-ph/9706344.

\bibitem{:2008jga}
A.~Airapetian et~al.
\newblock {Measurement of Azimuthal Asymmetries With Respect To Both Beam
  Charge and Transverse Target Polarization in Exclusive Electroproduction of
  Real Photons}.
\newblock {\em JHEP}, 06:066, 2008, 0802.2499.

\bibitem{Ackerstaff:1998av}
K.~Ackerstaff et~al.
\newblock {HERMES spectrometer}.
\newblock {\em Nucl. Instrum. Meth.}, A417:230--265, 1998, hep-ex/9806008.

\bibitem{Vanderhaeghen:1999xj}
M.~Vanderhaeghen, P.~A.~M. Guichon, and M.~Guidal.
\newblock {Deeply virtual electroproduction of photons and mesons on the
  nucleon: Leading order amplitudes and power corrections}.
\newblock {\em Phys. Rev.}, D60:094017, 1999, hep-ph/9905372.

\bibitem{Goeke:2001tz}
K.~Goeke, Maxim~V. Polyakov, and M.~Vanderhaeghen.
\newblock {Hard exclusive reactions and the structure of hadrons}.
\newblock {\em Prog. Part. Nucl. Phys.}, 47:401--515, 2001, hep-ph/0106012.

\bibitem{Polyakov:2002wz}
M.~V. Polyakov and A.~G. Shuvaev.
\newblock {On 'dual' parametrizations of generalized parton distributions}.
\newblock 2002, hep-ph/0207153.

\bibitem{Guzey:2006xi}
V.~Guzey and T.~Teckentrup.
\newblock {The dual parameterization of the proton generalized parton
  distribution functions H and E and description of the DVCS cross sections and
  asymmetries}.
\newblock {\em Phys. Rev.}, D74:054027, 2006, hep-ph/0607099.

\bibitem{Airapetian:2001yk}
A.~Airapetian et~al.
\newblock {Measurement of the beam spin azimuthal asymmetry associated with
  deeply-virtual Compton scattering}.
\newblock {\em Phys. Rev. Lett.}, 87:182001, 2001, hep-ex/0106068.

\bibitem{Airapetian:2006zr}
A.~Airapetian et~al.
\newblock {The beam-charge azimuthal asymmetry and deeply virtual Compton
  scattering}.
\newblock {\em Phys. Rev.}, D75:011103, 2007, hep-ex/0605108.

\end{thebibliography}

\end{footnotesize}

% ****************************************************************************
% END OF BIBLIOGRAPHY AREA
% ****************************************************************************

\end{document}